\documentclass[letterpaper]{article} 
\usepackage{aaai23}  
\usepackage{times}  
\usepackage{helvet}  
\usepackage{courier}  
\usepackage[hyphens]{url}  
\usepackage{graphicx} 
\usepackage{natbib}  
\usepackage{caption} 
\usepackage{subcaption}
\usepackage[inline]{enumitem}
\usepackage{booktabs}
\usepackage{amssymb}
\usepackage{multirow}
\newcommand{\ra}[1]{\renewcommand{\arraystretch}{#1}}

\usepackage{algorithm}
\usepackage{algorithmic}

\usepackage{newfloat}
\usepackage{listings}
\DeclareCaptionStyle{ruled}{labelfont=normalfont,labelsep=colon,strut=off} 
\floatstyle{ruled}
\newfloat{listing}{tb}{lst}{}
\floatname{listing}{Listing}
\title{Training Machine Learning Models to Characterize Temporal Evolution of Disadvantaged Communities}
\author{
    Milan Jain, 
    Narmadha Meenu Mohankumar, 
    Heng Wan, \\
    Sumitrra Ganguly, Kyle D Wilson, David M Anderson
}
\affiliations{
    Pacific Northwest National Laboratory, Richland, WA, USA\\
    \{milan.jain, narmadha.mohankumar, heng.wan, sumitrra.ganguli, kyle.wilson, dma\}@pnnl.gov
}

\usepackage{bibentry}

\begin{document}

\maketitle

\begin{abstract}
Disadvantaged communities (DAC), as defined by the Justice40 initiative of the Department of Energy (DOE), USA, identifies census tracts across the USA to determine where benefits of climate and energy investments are or are not currently accruing. The DAC status not only helps in determining the eligibility for future Justice40-related investments but is also critical for exploring ways to achieve equitable distribution of resources. However, designing inclusive and equitable strategies not just require a good understanding of current demographics, but also a deeper analysis of the transformations that happened in those demographics over the years. In this paper, machine learning (ML) models are trained on publicly available census data from recent years to classify the DAC status at the census tracts level and then the trained model is used to classify DAC status for historical years. 
A detailed analysis of the feature and model selection along with the evolution of disadvantaged communities between 2013 and 2018 is presented in this study.
\end{abstract}    
\section{Introduction}
In 2020, the Department of Energy (DOE) introduced Justice40 initiative, which directs 40\% of the overall benefits of certain Federal investments – including investments in clean energy and energy efficiency; clean transit; affordable and sustainable housing; training and workforce development; the remediation and reduction of legacy pollution; and the development of clean water infrastructure – to flow to disadvantaged communities (DACs)~\cite{dac_data}.
While using the percentile values of 36 indicators collected from numerous data sources, the initiative proposed a methodology to classify census tracts as Disadvantaged Communities (also referred to as DAC in the rest of the paper). The current version of DOE J40 DAC data (2022c) identifies 15,172 census tracts across the United States as DAC, out of which 262 belong to the state of WA (region of interest for this study).


A deeper understanding of these disadvantaged communities across the nation is crucial for designing equitable and inclusive policies for the communities. 
However, designing such inclusive and equitable strategies/policies not just requires a good understanding of current demographics, but also a deeper understanding of the transformations that happened in those demographics over the years. But the major hurdle in carrying out such a deep dive into the past is the data availability. The 36 indicator variables used by the Justice40 initiative 
are collected from numerous data sources, most of which were not even recorded in the past. 

Existing studies have have primarily focused on ensuring if the strategies are appropriately designed to respond to critical deficiencies in the DAC communities. Examples include digital-sharing economy, enterprise zone programs, and community development financial institutions and corporations for stimulating employment, income, reciprocity, social interaction, and resource accessibility for the DAC communities \cite{vidal1995reintegrating,dillahunt2015promise}. 
Akin to that, other studies have also argued that the disparity in spatial accessibility of infrastructure is strongly associated with inequalities among communities and that equitable distribution of public and private sector investments in infrastructure projects is critical \cite{leyshon1994access,brown2014spatial,mandarano2017equitable,wiesel2018locational}. While these studies exist, the studies that investigate the evolution of DAC status and the transformations of the determinants of DAC status over the years are important but lacking. 


In this study, we tackle this challenge by training Machine Learning (ML) models on different combinations of the LODES data (LEHD Origin-Destination Employment Statistics)~\cite{lodes_data} and the ACS (American Community Survey) data. For data description, please refer to the Appendix.  
The trained ML models act as a proxy for those 36 indicators in classifying the DAC status. Once trained, the best trained model is used to classify the DAC status of the census tracts for any time period for which the feature data is available. 
In this study, we particularly focused on feature selection and model selection to train most accurate model to project DAC for the historical data with limited bias. 

\section{Methodology}

\subsection{Data Collection \& Preprocessing}
For this study, we collected data from three sources: LEHD
Origin-Destination Employment Statistics (LODES) Data, American Community Survey (ACS) 5Y Estimates, and DOE Justice40 DAC Data. 
\begin{figure}[ht!]
    \centering
    \includegraphics[width=\columnwidth]{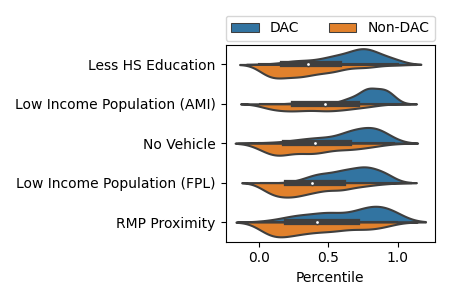}
    \caption{Top 5 indicators (out of 36 DOE J40 DAC indicators) best differentiate DACs from Non-DACs. \emph{Less HS Education} indicates the \% of total population, age $>$25, whose reported education is short of a high school diploma. \emph{Low-Income Population (AMI)} depicts the \% of the total population which is considered low income based on area median income (AMI) and \emph{Low-Income Population (FPL)} is the \% of the total population reported at or below 200\% of the Federal Poverty Level (FPL). \emph{RMP Proximity} indicates proximity to Risk Management Plan (RMP) facilities.}
    \label{fig:violin_dac}
\end{figure}
DOE Justice40 2022c edition of the data uses 2018 LODES data and 2019 ACS 5Y estimates. Figure~\ref{fig:violin_dac} shows the top 5 indicators (out of 36 DOE J40 DAC indicators) that best differentiates DACs from non-DACs. The percentile values shown on the x-axis are the rank of the specified indicator in the national data as a percent of the full dataset. The DAC score, based on which DAC status is decided, is the sum of those percentile values. 

It is evident from the plot that the features related to income and education are the most important. To capture these indicators directly/indirectly, the following five versions of training datasets were prepared for model training:
\begin{enumerate}
    \item \texttt{v1a: LODES(R)} - In this variant, we use all home-area characteristics (see Table~\ref{tab:lodes}) of the census tracts from the LODES data in the feature set. 
    \item \texttt{v1b: LODES(W)} - In this variant, we use all work-area characteristics (see Table~\ref{tab:lodes}) of the census tracts from the LODES data in the feature set.
    \item \texttt{v1c: LODES(R+W)} - In this variant, we use both home- and work-area characteristics (see Table~\ref{tab:lodes}) of the census tracts from the LODES data in the feature set.
    \item \texttt{v2a: LI(R)+ACS} - Though demographics (specifically race and ethnicity) are highly correlated with the DAC status, it is hard to intervene in demographics through policy design. In this variant, we exclude demographic information (age, sex, race, and ethnicity) and only incorporate the number of employed people and employment by industry from the LODES residential-area characteristics data. In addition, LODES income bins have low resolution, are not adjusted for inflation and do not capture household income. Since income is an important feature (as shown in Figure~\ref{fig:violin_dac}), high-resolution 16-bin household income adjusted for inflation from the ACS data is also included in the feature set. 
    \item \texttt{v2b: LI(R+W)+ACS} - The employment by industry from LODES residential-area characteristics only captures the distribution of industries where people work. To capture the distribution of industries that exist in a census tract, we incorporated employment by industry from LODES work-area characteristics (WAC) in this variant, along with existing features from the previous variant.
\end{enumerate}

Prior to training, the data is normalized with total number of employed people for \texttt{v1a: LODES(R)}, \texttt{v1b: LODES(W)}, and \texttt{v1c: LODES(R+W)}, and by total population for \texttt{v2a: LI(R)+ACS} and \texttt{v2b: LI(R+W)+ACS}.

\subsection{Model Training}
We used H2O.ai for model training and selection. H2O.AI is an AutoML framework~\cite{H2OAutoML20, H2OAutoML_33212} extensively used by the community for automating the ML workflow. Training included 30 different variants of 5 key models supported by the H2O.ai library: 
\begin{enumerate*}[label={(\arabic*)}]
    \item Distributed Random Forest (DRF),
    \item Deep Learning,
    \item Gradient Boosting Machines (GBM), 
    \item Generalized Linear Model (GLM),
    \item XGBoost, and
    \item Extremely Randomized Trees (XRT).
\end{enumerate*}
Model details can be found on H2O.AI algorithms page~\cite{H2OAutoML_33212}. 

For training, the data was split into 67:33, with 67\% data used to train 30 different models and remaining 33\% data for model evaluation and selection. The data was standardized by subtracting mean and dividing by standard deviation.

\subsection{Inference}
Though LODES data exists since 2002, a number of features were not included until 2009. Likewise, income information from ACS is only available starting 2013. Therefore, we used the best trained model to infer the DAC status for 5 years: 2013-2017. 
Lastly, we correlate temporal evolution of estimated DAC status with the most important features, as identified from the trained models. 

\section{Evaluation}

Table~\ref{tab:training_acc} depicts the F1-score of models trained on data from 968 census tracts (165 DACs), when evaluated on the test data which includes 477 census tracts (97 DACs) from the state of WA. Details about the tuned parameters of the best models are provided in Table~\ref{tab:hyperparameters} (on the last page). 

\begin{figure}[ht!]
    \centering
    \includegraphics[width=\columnwidth]{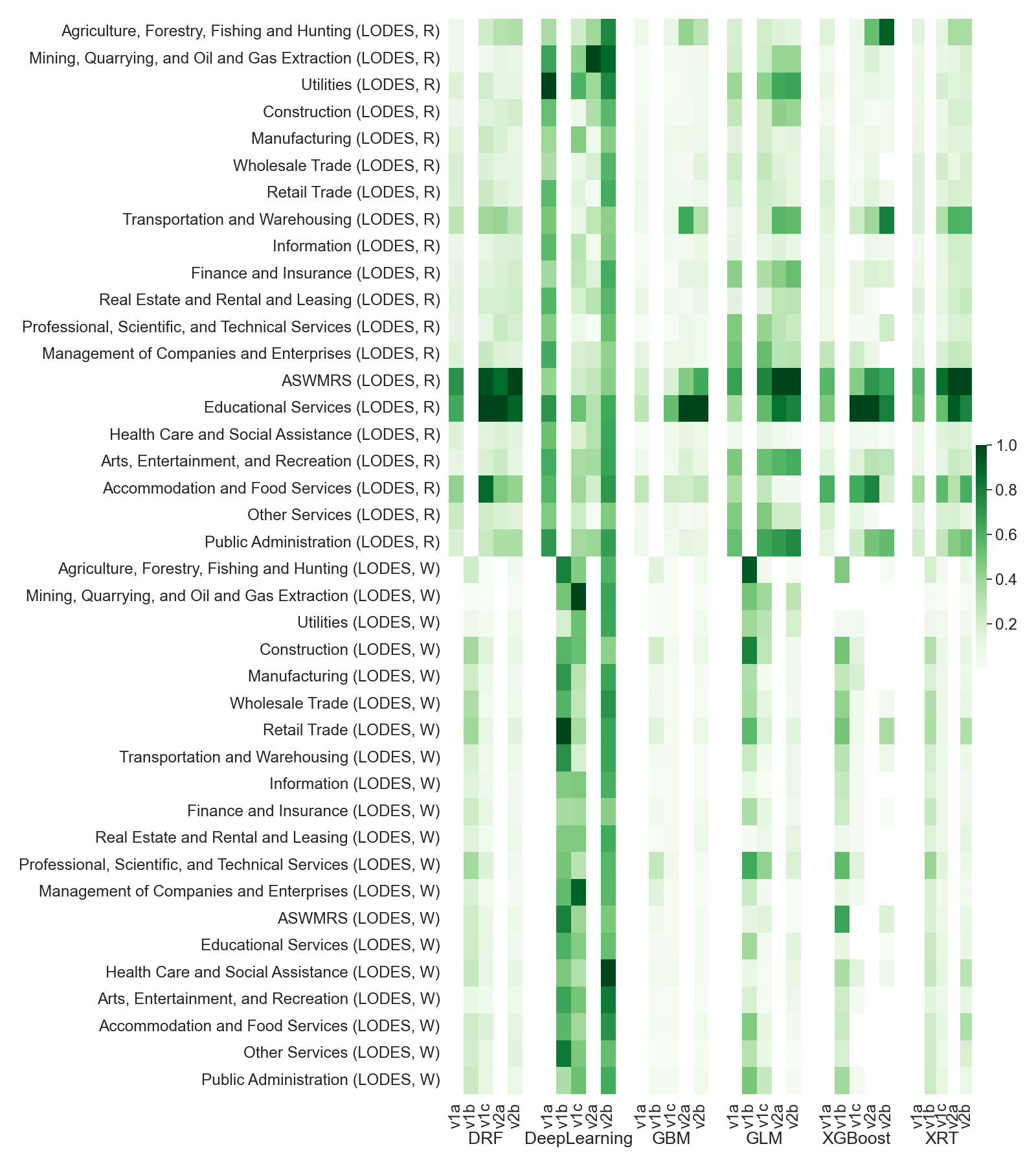}
    \caption{Feature Importance (Industry): x-axis shows data variants grouped by model. Here, feature importance only from the best variant of the model is shown.}
    \label{fig:fi_indstry}
\end{figure}

\begin{table}[ht!]
    \caption{Feature Engineering and Model Selection}
    \label{tab:training_acc}
    \ra{1.3}
    \resizebox{\columnwidth}{!}{\
    \begin{tabular}{lrrrrrr}
    \toprule
     &  DRF &  DeepLearning &  GBM &  GLM &  XGBoost &  XRT \\
    \midrule
    LODES(R) & \textbf{0.70} &  0.69 & 0.68 & 0.68 & 0.68 & 0.69 \\
    LODES(W)      & 0.49 &     0.50 & \textbf{0.55} & 0.52 &     0.53 & 0.52 \\
    LODES(R+W)    & 0.68 &     0.69 & 0.69 & 0.71 &     0.69 & \textbf{0.74} \\
    LI(R)+ACS   & 0.72 &          0.71 & 0.74 & 0.75 &     \textbf{0.75} & 0.69 \\
    LI(R+W)+ACS & 0.70 &          0.72 & \textbf{0.78} & 0.74 &     0.75 & 0.70 \\
    \bottomrule
    \end{tabular}}
\end{table}
\begin{figure}[ht!]
    \centering
    \includegraphics[width=\columnwidth]{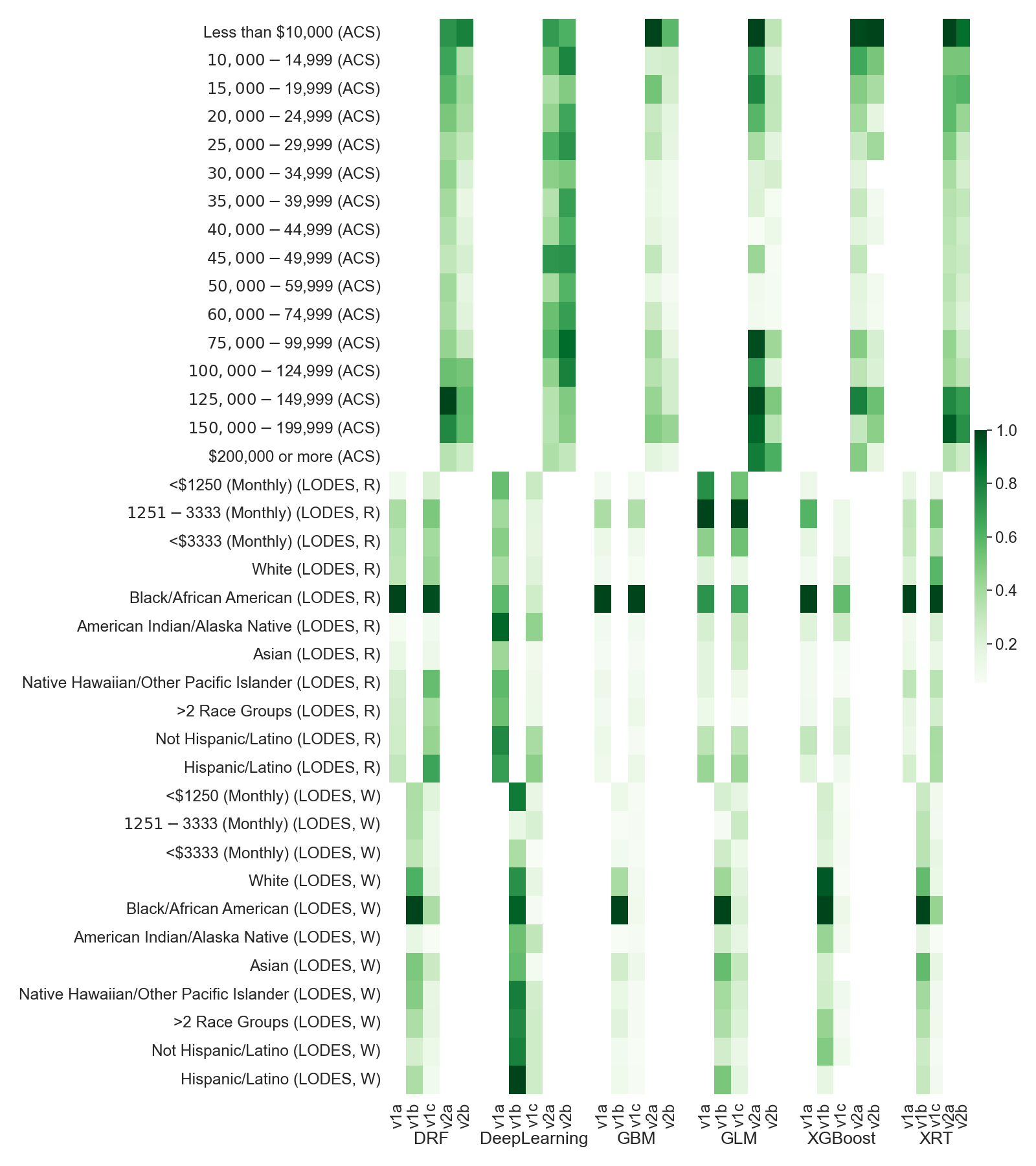}
    \caption{Feature Importance (Race, Ethnicity, and Income): x-axis shows data variants grouped by model. Here, feature importance only from the best variant of the model is shown.}
    \label{fig:fi_rei}
\end{figure}
Figure~\ref{fig:fi_rei} shows the importance of features related to race, ethnicity, and income, and Figure~\ref{fig:fi_indstry} shows the feature importance of employment by industry features for every combination of data variant and ML model (its best version). For GLM, variable importance indicates the coefficient magnitudes. For tree based algorithms (GBM, DRF, XRT, and XGBoost), the variable importance is determined by calculating the relative influence of each variable: whether that variable was selected to split on during the tree building process, and how much the squared error (over all trees) improved (decreased) as a result. Finally, for the Deep Learning model, H2O.AI uses Gedeon method, that considers the weights connecting the input features to the first two hidden layers to compute the variable importance~\cite{gedeon1997data}.

\begin{figure}[t!]
    \centering
    \begin{subfigure}[b]{0.49\columnwidth}
        \centering
        \includegraphics[width=\columnwidth, trim={2cm 2cm 2cm 2.5cm}, clip]{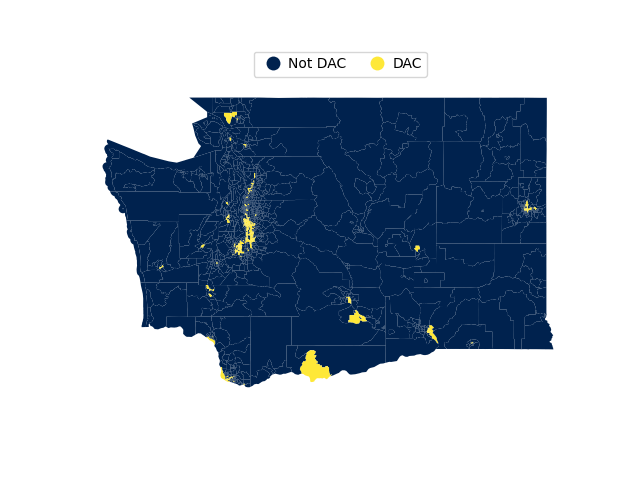}
        \caption{Actual}
        \label{fig:wa_act}
    \end{subfigure}
    \begin{subfigure}[b]{0.49\columnwidth}
        \centering
        \includegraphics[width=\columnwidth, trim={2cm 2cm 2cm 2.5cm}, clip]{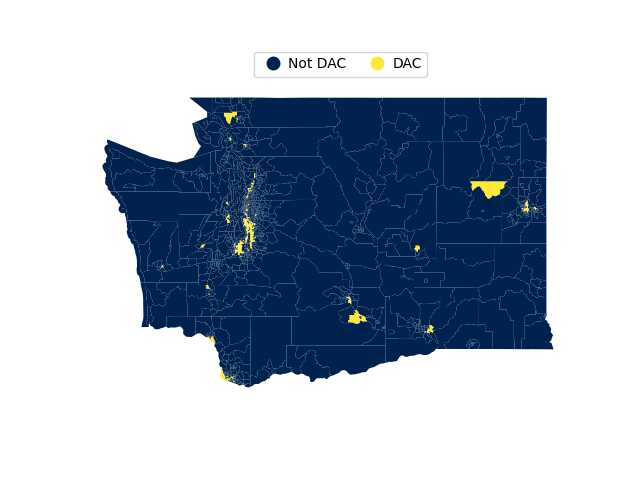}
        \caption{Predicted}
        \label{fig:wa_pred}
    \end{subfigure}
    \caption{Comparing Actual DAC communities with the predicted DAC communities.}
    \label{fig:pred_comp}
\end{figure}
\begin{figure*}
    \centering
    \includegraphics[width=\textwidth, trim={1cm 5.5cm 0 4cm}, clip]{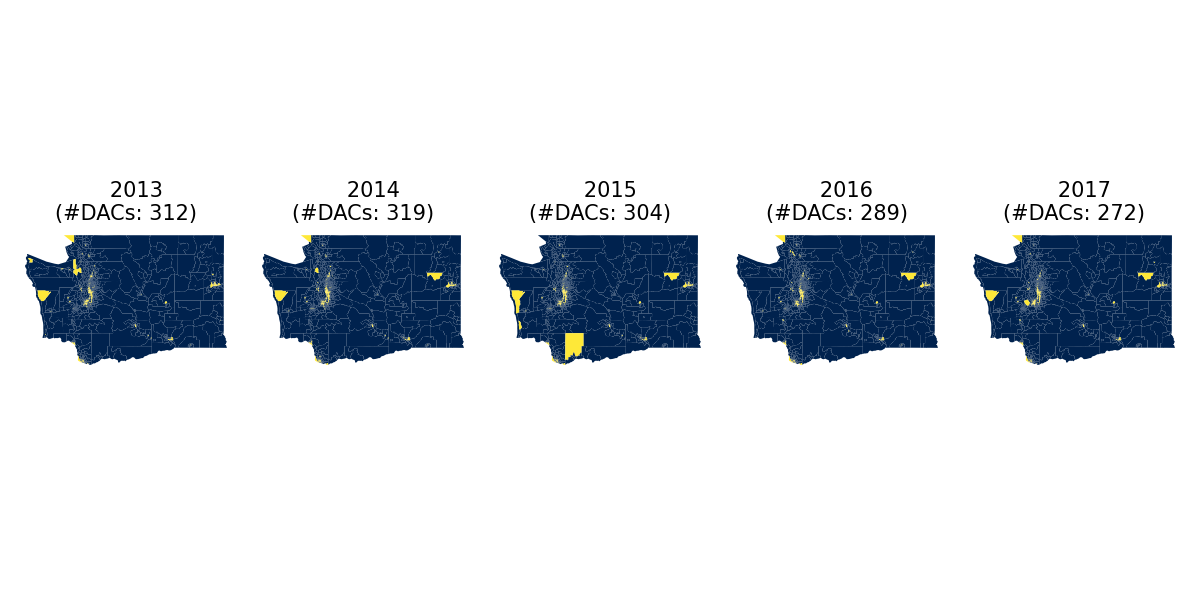}
    \caption{DAC Estimation on Historical Data (2013-2017)}
    \label{fig:historical}
\end{figure*}

Following are some key takeaways:
\begin{itemize}
    \item Residential-area characteristics alone are better estimator of DAC status than the work-area characteristics. 
    \item Combining residential- and work- area characteristics offer very little improvement over residential-area characteristics alone (comparing \texttt{v1a: LODES(R)} with \texttt{v1c: LODES(R+W)}).
    \item Models trained on \texttt{v1a: LODES(R)}, \texttt{v1b: LODES(W)}, and \texttt{v1c: LODES(R+W)} features rely heavily on race and ethnicity distributions (from Figure~\ref{fig:fi_rei}). Dependence on race and ethnicity introduces bias in the model. For instance, the biased model trained on these features sets was projecting census tracts with high African-American population as DAC.
    \item Removing demographic information (race, ethnicity, gender, and age) and only using employment by industry from LODES residential area-characteristics along with high-resolution income bins from ACS (\texttt{LI(R)+ACS}) improved the DAC estimation accuracy.
    \item Three bin income categorization from LODES only covered income of employed people and hardly provided any separation between DAC/non-DAC to the model. Instead, 16-bins of household income (adjusted for inflation) from ACS seems like better feature set for the DAC classification, especially the low- and high-income bins (except for Deep Learning), as shown in  Figure~\ref{fig:fi_rei}.  
    \item Industries like Transportation and Warehousing, Educational Services, Administrative and Support and Waste Management and Remediation Services (ASWMRS), Accommodation and Food Services, and Public Administration come up as important features (Figure~\ref{fig:fi_indstry}). 
    \item Employment by industry from residential-area characteristics only capture \emph{where people work} and not \emph{what kind of industries operate in the area}.
    To address this, we incorporated employment by industry from work-area characteristics in \texttt{LI(R+W)+ACS}. However, including this additional information into the feature set offers very limited improvement, if any, across all the models.
\end{itemize}

It is evident from the analysis that Gradient Boosting Machine (GBM) using \texttt{v2b: LI(R+W)+ACS} feature set provides the best DAC classification accuracy of 78\%, without any bias towards a particular community. Figure~\ref{fig:pred_comp} compares actual DACs with the estimated DACs on the map of WA state. While 89\% census tracts are correctly identified, there are a few false positives and a few false negatives.


\subsubsection{False Negatives (what model missed!):}
On analyzing false negatives, we noticed that the census tracts that model failed to identify as DAC failed on two key indicators of DOE J40 DAC definition - Low Income Population (AMI) and Low Income Population (FPL). 
When compared with true positive tracts, low income related indicators are relatively lower for false negative tracts. The trained model missed these tracts because income is a an important feature (see Figure~\ref{fig:violin_dac}). A natural follow-up question is - \emph{why these tracts are DACs then?} Our analysis found that the age of the house (exposure to lead) is a key factor driving the DAC status of these tracts, which is not captured in the feature set.   

\subsubsection{False Positives (model thinks it's a DAC!):}
On analyzing the false-positives, we noticed that a number of false positive instances are those census tracts that belong to big cities like Seattle and Spokane. Typically, in such census tracts, industries like Transportation, Waste Management, and Public Services have substantial presence - an important set features for DAC estimation by the trained models, as shown in Figure~\ref{fig:fi_indstry}. Besides, these tracts have relatively high percentage of households from both low and high income groups, which is another set of important features of the trained models, as shown in Figure~\ref{fig:fi_rei}. However, these tracts were not identified as a DAC by the DOE J40 DAC definition.

Though the false positive rate can be reduced by incorporating additional features differentiating big cities from relatively smaller cities, they do offer an opportunity to further analyze those communities as potential DAC communities. Given that the DAC definition is still in experimental stage, evaluation of such false positive instances as potential DAC community becomes even more important. 


\subsection{DAC Estimation (Historical Data)}
Overall, though the disadvantaged census tracts is distributed very similarly between 2013-2017, the total number of DAC communities seems to have decreased over time. When correlated with important features (from Figure~\ref{fig:fi_rei} \& \ref{fig:fi_indstry}), a decrease in low-income (household) group and increase in high-income (household) group were noticed in the state of WA between 2013 and 2017.
One must note here that the household income reported in ACS is already adjusted for inflation. Since income is an importance feature, decrease in low-income group and increase in high-income group explains the reduction in number of DAC communities between 2013 and 2017. 
However, these correlations doesn’t imply causation and a deeper analysis is required to identify true causes. Once identified, those causal links would assist the stakeholders and the decision makers in designing equitable and inclusive policies.
\section{Conclusion}
Designing inclusive and equitable strategies not just requires a good understanding of current demographics, but also a deep dive into the transformations that happened in those demographics over years. In this paper, we used AutoML to train several machine learning (ML) models on LODES and ACS data to classify the DAC status at the census tracts level and used the best trained model to classify DAC status between 2013-2017. 
Our analysis indicates that the Gradient Boosting Machine on features related to employment and income is the most accurate model with no bias towards any community. When used on historical data, we noticed a decline in number of disadvantaged communities between 2013 and 2017. The decline seems to be correlated with reduction in low-income groups and increase in high-income groups, some of the most important features of the trained models. 
However, one must note here that these correlations doesn't imply causation and further analysis is required to identify the true causes.
\section{Acknowledgments}

This research was supported by the Agile Initiative, a multi-disciplinary Pacific Northwest National Laboratory (PNNL) initiative. PNNL is operated by Battelle Memorial Institute under Contract DE-
AC06-76RL01830.

\bibliography{ref}

\begin{table}[!ht]
    \centering
    \begin{tabular}{|p{1.5cm}|p{3.8cm}|c|c|}
    \hline
    Parameter & Categories & RAC & WAC \\ [0.5ex] 
    \hline\hline
    Age (in Years) &  $\leq$29; 30-54; 55$\geq$ & $\checkmark$ & $\checkmark$\\ \hline
    Income (in \$/Month) & $\leq$1250; 1250-3333; 3333$\geq$ & $\checkmark$ & $\checkmark$ \\ \hline
    Industry & 20 categories\footnote{Please refer to LODES documentation for the detailed list.} & $\checkmark$ & $\checkmark$ \\ \hline
    Race & White; African American; American Indian or Alaska Native; Asian; Native Hawaiian or Other Pacific Islander; Two or More Race Groups & $\checkmark$ & $\checkmark$  \\ \hline
    Ethnicity &  Not Hispanic or Latino; Hispanic or Latino & $\checkmark$ & $\checkmark$\\ \hline
    Education &  Less than high school; High school or equivalent, no college; Some college or associate degree; Bachelor's degree or advanced degree & $\checkmark$ & $\checkmark$\\ \hline
    Gender &  Male; Female & $\checkmark$ & $\checkmark$\\ \hline
    Firm Age (in Years) &  0-1; 2-3; 4-5; 6-10; 11+ &  & $\checkmark$\\ \hline
    Firm Size &  0-19; 20-49; 50-249; 250-499; 500+ &  & $\checkmark$\\ \hline
    \end{tabular}
    \caption{LEHD Origin-Destination Employment Statistics (LODES) Data}
    \label{tab:lodes}
\end{table}
\begin{table}[!ht]
    \centering
    \ra{1.3}
    \begin{tabular}{|l|p{2.4cm}|}
    \hline
    Parameter & Categories \\ [0.5ex] 
    \hline\hline
    Household Income (Annual) & $\leq$10000; 10000-14999; 15000-19999; 20000-24999; 25000-29999; 30000-34999; 35000-39999; 40000-44999; 45000-49999; 50000-54999; 55000-59999; 60000-74999; 75000-99999; 100000-124999; 125000-149999; 150000-199999; $\geq$200000 \\
    \hline
    \end{tabular}
    \caption{American Community Survey (ACS)}
    \label{tab:acs}
\end{table}
\appendix
\section{Data Description}
\label{app:data_desc}

\paragraph{LODES Data:}
Published by the U.S. Census Bureau, the LODES data~\cite{lodes_data} primarily captures the employment statistics at the block-level by demographics and industries. It is organized into three groups: (1) OD – Origin-Destination data associated with transition of employed population between home and work census blocks, (2) RAC – Residence Area Characteristics data by home census block i.e. distribution of people that live there, and (3) WAC – Workplace Area Characteristics data by work census block i.e. distribution of people that work there. Table~\ref{tab:lodes} shows all the demographics features and their corresponding categories available in the LODES data. RAC and WAC columns indicate the availability of a feature in either dataset.

\paragraph{ACS Data:}
The American Community Survey (ACS)~\cite{ACS} is a demographics survey program conducted by the U.S. Census Bureau every year and covers a broad range of topics about social, economic, demographic, and housing characteristics of the U.S. population. The 5Y estimates summarize sample data collected from last five years at the block-group level. Over 1Y estimates, 5Y estimates provide increased statistical reliability of the data for less populated areas and small population subgroups. For the purpose of this study, we only used 16-bin household income, adjusted for inflation, at the block-group level from the ACS data (see Table~\ref{tab:acs}). 

\paragraph{DAC Data:}
DAC~\cite{dac_data} is the U.S. Department of Energy's working definition of disadvantaged communities as pertaining to \textsc{EO 14008}, or the Justice40 Initiative. The DAC data includes thirty-six (36) burden indicators collected at the census tract level and an indicator identifying each census tract as disadvantaged or not disadvantaged. The 36 indicators are taken from various data sources including American Community Survey (ACS), Longitudinal Employer-Household Dynamics (LEHD) Survey, Environmental Justice Screening Tool (EJScreen), among others. For this study, we are using 2022c version of the DAC data. For detailed information, please refer to Justice40 DAC data documentation~\cite{dac_data}.

\section{Model Hyperparameters}
Table~\ref{tab:hyperparameters} provides details about the tuned parameters of the best models from H2O.AI.
\begin{table*}[ht!]
    \centering
    \caption{Best Hyperparameters from Grid Search}
    \label{tab:hyperparameters}
    \ra{1.3}
    \resizebox{\textwidth}{!}{\
    \begin{tabular}{lrrrrr}
    \toprule
    {} & v1a:LODES(R) & v1b:LODES(W) &        v1c:LODES(R+W) & v2a:LI(R)+ACS & v2b:LI(R+W)+ACS\\
    \midrule
    \textbf{Gradient Boosting Machine (GBM)}\\
    col\_sample\_rate                  &           0.7 &           0.8 &              0.8 &              1.0 &                0.4 \\
    col\_sample\_rate\_per\_tree         &           1.0 &           0.8 &              0.8 &              1.0 &                0.7 \\
    learn\_rate                       &           0.1 &           0.1 &              0.1 &              0.1 &                0.1 \\
    max\_depth                        &             4 &            15 &                7 &               17 &                  6 \\
    min\_rows                         &           5.0 &         100.0 &             10.0 &             15.0 &               10.0 \\
    min\_split\_improvement            &       0.00001 &       0.00001 &          0.00001 &          0.00001 &            0.00001 \\
    ntrees                           &            35 &            41 &               37 &               45 &                 50 \\
    sample\_rate                      &           0.9 &           0.8 &              0.8 &              0.9 &                0.5 \\
    \textbf{XGBoost}\\
    booster                          &        gbtree &        gbtree &           gbtree &           gbtree &             gbtree \\
    col\_sample\_rate                  &           0.8 &           0.8 &              0.8 &              0.8 &                0.6 \\
    col\_sample\_rate\_per\_tree         &           0.8 &           0.8 &              0.8 &              0.7 &                0.8 \\
    max\_depth                        &             5 &             5 &               10 &                9 &                  9 \\
    min\_rows                         &           3.0 &           3.0 &              5.0 &              5.0 &               10.0 \\
    ntrees                           &            34 &            33 &               35 &               42 &                 40 \\
    reg\_alpha                        &           0.0 &           0.0 &              0.0 &              1.0 &              0.001 \\
    reg\_lambda                       &           1.0 &           1.0 &              1.0 &              1.0 &               0.01 \\
    sample\_rate                      &           0.8 &           0.8 &              0.6 &              0.6 &                0.6 \\
    \textbf{Generalized Linear Model (GLM)}\\
    alpha                            &         [0.0] &         [0.0] &            [0.0] &            [0.0] &              [0.0] \\
    \textbf{Deep Learning}\\
    epsilon                          &           0.0 &           0.0 &         0.000001 &         0.000001 &                0.0 \\
    hidden                           &    [100, 100] &  [10, 10, 10] &     [50, 50, 50] &             [50] &              [100] \\
    hidden\_dropout\_ratios            &    [0.1, 0.1] &          None &  [0.4, 0.4, 0.4] &            [0.4] &              [0.1] \\
    input\_dropout\_ratio              &          0.15 &           0.0 &              0.2 &              0.2 &               0.15 \\
    rho                              &           0.9 &          0.99 &             0.95 &             0.95 &                0.9 \\
    \textbf{Distributed Random Forest (DRF)}\\
    balance\_classes                  &         False &         False &            False &            False &              False \\
    ntrees                           &            34 &            41 &               40 &               33 &                 33 \\
    max\_depth                        &            20 &            20 &               20 &               20 &                 20 \\
    col\_sample\_rate\_change\_per\_level &           1.0 &           1.0 &              1.0 &              1.0 &                1.0 \\
    col\_sample\_rate\_per\_tree         &           1.0 &           1.0 &              1.0 &              1.0 &                1.0  \\
    min\_split\_improvement            &       0.00001 &       0.00001 &          0.00001 &          0.00001 &            0.00001 \\
    \textbf{Extreme Random Forest (XRT)}\\
    balance\_classes                  &         False &         False &            False &            False &              False \\
    ntrees                           &            43 &            45 &               35 &               43 &                 41 \\
    max\_depth                        &            20 &            20 &               20 &               20 &                 20 \\
    col\_sample\_rate\_change\_per\_level &           1.0 &           1.0 &              1.0 &              1.0 &                1.0 \\
    col\_sample\_rate\_per\_tree         &           1.0 &           1.0 &              1.0 &              1.0 &                1.0 \\
    min\_split\_improvement            &       0.00001 &       0.00001 &          0.00001 &          0.00001 &            0.00001 \\
    \bottomrule
    \end{tabular}}
\end{table*}

\end{document}